# Anisotropic Electronic Structure of the Two-Dimensional Electron Gas at the AlO$_x$/KTaO$_3$(110) interface

E. A. Martínez[a], J. Dai[b], M. Tallarida[b], N. M. Nemes[a] and F. Y. Bruno[a]

[a]*GFMC, Departamento de Física de Materiales, Universidad Complutense de Madrid, 28040, Madrid, Spain*

[b]*ALBA Synchrotron Light Source, 08290, Cerdanyola del Vallès, Barcelona, Spain*

Corresponding Author: Flavio Y. Bruno, fybruno@ucm.es

**Abstract**

Oxide-based two-dimensional electron gases (2DEGs) have generated significant interest due to their potential for discovering novel physical properties. Among these, 2DEGs formed in KTaO$_3$ stand out due to the recently discovered crystal face-dependent superconductivity and large Rashba splitting, both of which hold potential for future oxide electronics devices. In this work, angle-resolved photoemission spectroscopy is used to study the electronic structure of the 2DEG formed at the (110) surface of KTaO$_3$ after deposition of a thin Al layer. Our experiments revealed a remarkable anisotropy in the orbital character of the electron-like dispersive bands, which form a Fermi surface consisting of two elliptical contours with their major axes perpendicular to each other. The measured electronic structure is used to constrain the modeling parameters of self-consistent tight-binding slab calculations of the band structure. In these calculations, an anisotropic Rashba splitting is found with a value as large as 4 meV at the Fermi level along the [-110] crystallographic direction. This large unconventional and anisotropic Rashba splitting is rationalized based on the orbital angular momentum formulation. These findings provide insights into the interpretation of spin-orbitronics experiments and help to constrain models for superconductivity in the KTO(110)-2DEG system.

**Introduction**

The bulk physical properties of transition metal oxides (TMOs) are determined by strongly correlated *d* electrons. In these materials, the interplay between lattice, charge, spin and orbital degrees of freedom gives rise to a variety of phenomena, such as superconductivity, magnetism, ferroelectricity, and charge and orbital ordering.[1] The interfaces of TMOs offer an interesting platform for exploring new properties, as the broken spatial inversion symmetry can be tailored, and the two-dimensionality enhances the influence of electron correlations.[2] The discovery of a two-dimensional electron gas (2DEG) at the interface between the TMOs LaAlO$_3$ (LAO) and SrTiO$_3$ (STO) spurred a large interest and soon a range of STO-based heterostructures with many fascinating properties emerged.[3,4]. Eventually, other 2DEGs were found in similar TMOs, for instance, KTaO$_3$ (KTO). The vast majority of studies were focused on 2DEGs confined along the [001] direction of the perovskite structure, which is common for both STO and KTO. In these systems, the free carriers move parallel to the (001) surface of the crystal and will be labeled STO(001) or KTO(001) in this work. However, it was later realized that the conducting system can also be formed starting from (110) and (111) surfaces, and the physical properties of the different systems were found to be dependent on the crystallographic orientation, highlighting the importance of the confinement direction.[5,6]

The discovery of superconductivity in the KTO(001)-2DEG some years ago generated a significant level of excitement as this phase was never observed in the bulk-doped material. However, the low critical temperature ($T_c$) ~ 50 mK tempered the enthusiasm to some extent. [7] This situation radically changed with the recent observation of superconductivity in the KTO(111) and (110) 2DEGs, as the reported $T_c$'s have reached 2.2 K and 0.9 K, respectively.[8,9] The pairing mechanism is still under discussion but electron-phonon coupling mediated by inter-orbital interactions appear to play a major role.[10] The KTO based superconducting 2DEGs were formed by depositing either amorphous LaAlO$_3$ (LAO) or EuO on top of KTO, and it was recently demonstrated that superconductivity is also observed at the surface of KTO gated with ionic liquid with similar $T_c$'s.[6] These results indicate that the physics of the problem is mostly related to the KTO surface orientation and the interplay among the different degrees of freedom at the interface, whereas the materials deposited on top of KTO or the ionic liquid play a role as a charge reservoir. A second salient feature of KTO-based 2DEGs is its potential in spin-orbitronics applications.[11,12] The large spin-orbit coupling of Ta $5d$ atoms, and the broken spatial inversion symmetry found at the surface and/or interface, are the necessary ingredients for the lifting of the spin degeneracy in the 2DEG, broadly known as the Rashba effect.[13,14] Through the so-called direct and inverse Edelstein effects, these 2DEGs can be used to generate spin currents from charge currents and vice versa.[11,15–17] The spin-charge interconversion is closely related to the band structure of the 2DEG, and a larger Rashba spin-splitting would enhance the conversion efficiency.[11,18] In this context, the determination of the electronic band structure of KTO-based 2DEGs will constitute a step towards understanding both superconductivity and Rashba spin-splitting in the system.

Angle-resolved photoemission spectroscopy (ARPES) has been used extensively for the direct determination of the electronic structure of TMOs-based 2DEGs.[19,20] Direct visualization of characteristics such as Fermi surface volume and subband dependent orbital character in ARPES measurements provide constraints on the modeling parameters allowing for more realistic calculations, thereby helping to elucidate the complex spin and orbital textures of the band structure.[21–25] These experiments are feasible on the bare surface of oxides, where 2DEGs are formed by the introduction of oxygen vacancies created by irradiation with photons under UHV conditions or, alternatively, by depositing a thin Al layer that pumps oxygen from the substrate in an efficient redox reaction.[26,27] Probing the surface of different TMOs, it has been shown that it is possible to stabilize a 2DEG in STO, KTO, TiO$_2$, CaTiO$_3$, among others.[19,28–35] Considering that ARPES is a surface sensitive technique, and the 2DEG resides in the first few layers of the crystal, high quality measurements require a clean and ordered surface. Materials like KTO pose a problem from an experimental point of view, since methods for preparing a clean and atomically ordered surface are cumbersome, and only exist for certain crystallographic orientations.[36,37] High quality ARPES results have been obtained for the 2DEG stabilized in the KTO(001) and (111) surfaces.[30,31,38,39] In the bulk band structure of undoped, insulating KTO, the empty conduction bands are mainly derived from the $t_{2g}$ manifold: $d_{xy}$, $d_{yz}$ and $d_{zx}$ orbitals. These bands break their orbital degeneracy at the Γ point due to the intrinsically strong spin-orbit (SO) interaction. In consequence, two pairs of spin degenerate $J = 3/2$ bands define the conduction band minimum (CBM), and an additional spin degenerate band with $J = 1/2$ is placed 400 meV above the CBM at the Γ point. In both KTO(001) and (111), it was found that the 2DEG is formed by the $J = 3/2$ states without contribution from $J = 1/2$ states. Importantly, confinement along the [001] direction results in a breaking of the degeneracy at the Γ point, as a consequence of the real-space anisotropy of the orbital wavefunctions along the confinement direction.[21,30,31] On the contrary, $t_{2g}$ orbitals are equivalent except for a rotation of 120º along the [111] confinement direction, hence no induced orbital polarization is expected.[38] While ARPES was used to gain understanding of the role of confinement along the [110] direction in STO, the situation for the $J = 3/2$ bands of KTO remains to be explored.[28,40]

In this communication, we study the electronic structure of the 2DEG formed in KTO(110) by deposition of a thin Al layer on the crystal surface. Our XPS measurements reveal the existence of electrons doped into KTO and an additional spectral weight at the Fermi level after Al deposition. We use ARPES to study these electron-like dispersive bands and find a Fermi surface composed of two elliptical sheets with their major axes perpendicular to each other and different orbital character. The Fermi wavevectors obtained from the ARPES measurements, related to the 2DEG carrier density ($n_{2D}$), are used to constrain our electronic structure calculations. The computed band structure is in fair agreement with the measured orbital characters and bandwidth of the electronic system and allows us to study details of the electronic structure not resolved in the experiment. In particular, we identify an unconventional Rashba-like lifting of the spin degeneracy with distinctive features: firstly, it is larger than in any other previously studied STO- or KTO-based 2DEG, and secondly, it is not originated in the avoided crossing points of the bands but related with an enhanced interplay between orbital and spin angular momentum.

**Results and Discussion**

In order to stabilize the 2DEG, we used KTO(110) substrates from two different providers, SurfaceNet and PiKEM, with no significant differences observed in our experiments between them. The substrates were *in-situ* annealed at 500 ºC for 30 minutes in a vacuum better than $10^{-7}$ mbar to ensure the cleanliness of the surface. No further surface treatment was performed. Subsequently, we deposited 1 Å of Al with a calibrated evaporation source at room temperature and a pressure P = $10^{-8}$ mbar and immediately transferred the samples for ARPES measurements. After deposition of Al on the surface of KTO, the Al oxidizes into AlO$_x$ pumping oxygen from the KTO substrates, the resulting positively charged oxygen vacancies (OVs) stay close to the Al/KTO interface and act as electron donors to the system.[27] The mobile electrons will partially fill the *5d* bands of KTO and try to screen the positively charged OVs creating the 2DEG. We verify the presence of electron doping in KTO by measuring the XPS spectra corresponding to Al *2p* and Ta *4f* as shown in Figure 1(a) and (b), respectively. The presence of a peak at ~ 74.4 eV binding energy (BE) corresponds to Al$^{3+}$ in AlO$_x$, while there is negligible intensity at ~72.5 eV that would correspond to metallic Al$^0$. In the Figure 1(b), we can see the doublet that corresponds to the *4f* levels of Ta$^{5+}$. The peaks at ~26.2 eV and 28.1 eV are the Ta *4f$_{7/2}$* and *4f$_{5/2}$* components of the Ta 4f spin-orbit doublet. The presence of a shoulder at ~24.8 eV in the Ta *4f* spectrum indicates that the reduced Ta$^{4+}$ specie is present and thus, there are carriers filling the Ta-*5d* conduction bands. The relative fraction $I_{Ta4+}/I_{Ta5+} \sim 0.04$, obtained by fitting the spectrum with both oxidation states components, is consistent with previous observations and results in the presence of a 2DEG in KTO(001).[41,42]. Looking at the valence band in Figure 1(c), we found a small increase in spectral weight at the Fermi level ($E_F$) which corresponds to the conductive states of the 2DEG. Together with the rise of the 2DEG peak, there is an increase in spectral weight with BE ~1.5 eV, this bump of in-gap (IG) states is commonly associated with oxygen vacancies or other defects.[30,31,43,44]. Importantly, the oxidized Al layer is insulating and does not contribute spectral weight near the $E_F$. Using a thicker Al layer results in the presence of a metallic Al$^0$ peak and a large nondispersive intensity at the $E_F$, which precludes ARPES measurements of the 2DEG.[27,45]

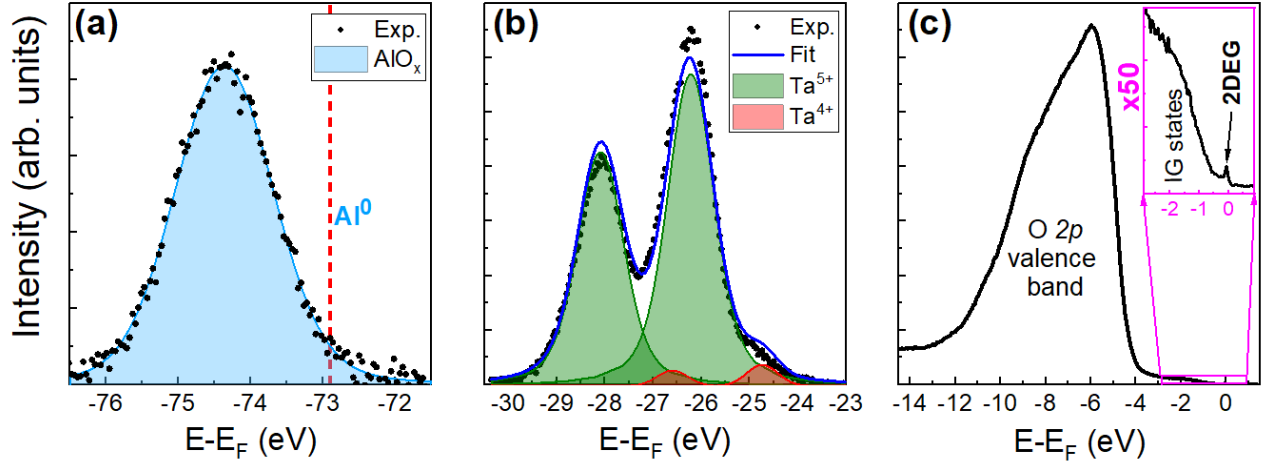

**Figure 1.** (a) XPS spectra of Al *2p*, (b) Ta *4f* core levels and the respective fits measured with a photon energy of 110 eV. No signal of Al$^0$ is detected in the Al *2p* spectrum, while in the Ta *4f* spectrum, a small shoulder indicates the presence of Ta$^{4+}$. (c) The valence band dominated by O *2p* states is shown. (Inset) close-up view of the spectra revealing the presence of a peak at the Fermi level originated by the 2DEG. The valence band measurements were obtained using a photon energy of 65 eV.

In Figure 2(a) and 2(b) we present energy-momentum measurements of the 2DEG along the X-Γ-X path, *i. e.*, the [-110] direction, obtained using a photon energy of $h\nu = 65$ eV with linear horizontal (LH) and vertical (LV) beam polarizations, respectively. The experimental geometry for the ARPES measurement is depicted in the schematic of Figure 2(e) where the detection and sample plane are shown together with the beam electric field (**E**) for each polarization. An electron-like band with a ~120 meV bandwidth and Fermi wavevector $k_F = 0.16$ Å$^{-1}$ is observed with LV photons, while a narrower ~100 meV bandwidth band with $k_F = 0.06$ Å$^{-1}$ is detected with LH light. The linewidth of the spectra is ~ 0.1 Å$^{-1}$, too broad to inspect fine details of the band structure and much broader than expected from the spectral resolution of the experimental setup.[42] Fermi surface measurements taken with LH and LV beam polarization are shown in Figure 2(c) and 2(d), respectively. Two distinct elliptical shapes, with their major axes perpendicular to each other, are easily recognizable. In Figure 2(f) a schematic of the 2D first Brillouin zone (BZ1) of KTO(110) along with a size comparison with the Fermi contours is shown. As observed in other 2DEGs, these Fermi surface contours are different to the contours that would be found by taking a 2D cut in the (110) plane of the bulk electronic structure. Instead, they are rather similar to the projection of the bulk Fermi surface onto the (110) plane. [42,46] The dependence of the measurements on the beam polarization is related to the orbital character of the electronic structure and the experimental setup geometry. The symmetry of the matrix elements can be understood by considering the symmetry of the 5*d* $t_{2g}$ electrons with respect to the crystallographic orientation.[42,47] Figure 2(g) shows the plots of the Ta $t_{2g}$ maximally localized Wannier functions (MLWFs) representing the 5$d_{xy}$, $d_{yz}$ and $d_{zx}$ orbitals, as seen from the [110] direction. [48] By examining the parity of the wavefunctions relative to the sample mirror plane and the detection plane, we can determine the detected orbital characters in the photoemission process. In our experiment, when LH (LV) photons are incident on the surface along the [-110] direction, we observe $d_{xy}$ ($d_{zx}/d_{yz}$) orbital characters preferentially. As a result, the states seen in Figure 2(a) and (c) exhibit mostly $d_{xy}$ orbital character, while those in Figure 2(b) and (d) display a mixture of $d_{zx}/d_{yz}$ orbital characters. The experimentally determined Fermi wavevectors, bandwidth and the orbital character for the electronic structure of KTO(110)-2DEG are used to constrain and validate our electronic structure calculations, which in turn allow us to inspect details not observed in the experiment.

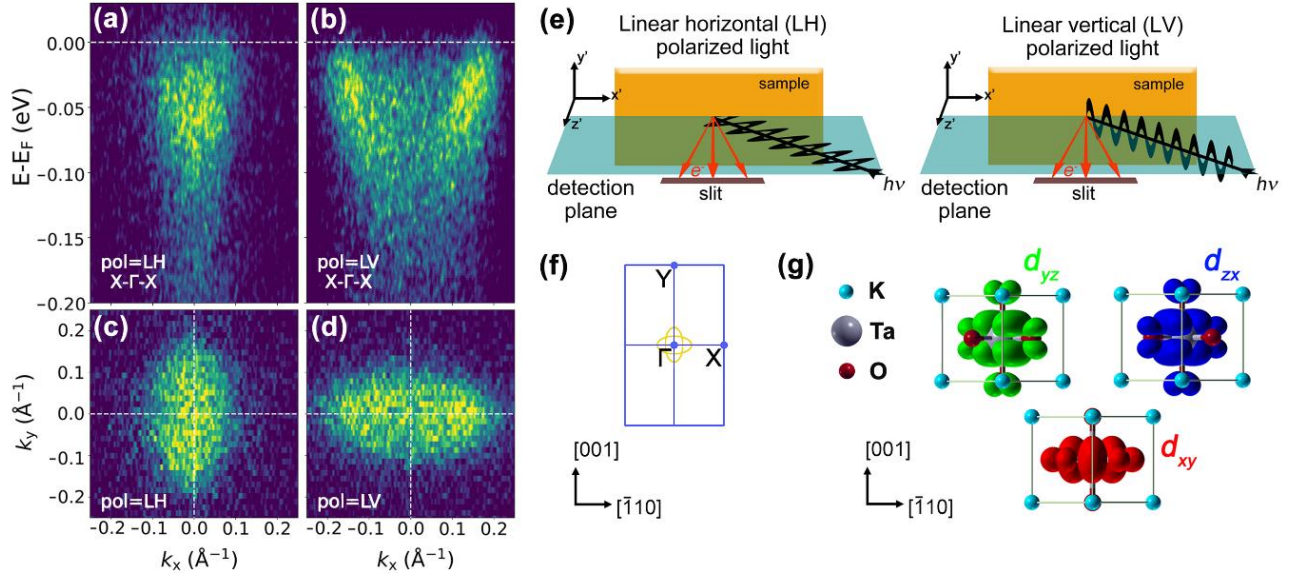

**Figure 2.** (a, b) Energy-momentum dispersions measured by ARPES along X-Γ-X with LH and LV beam polarizations. (c, d) Fermi surface measured with LH and LV beam polarizations. (e) Schematics of the ARPES experimental setup for the LH and LV polarizations as found in ALBA-LOREA beamline. (f) First Brillouin zone (BZ1) and crystallographic orientations. The contours at the Γ-point correspond to the relative size of the Fermi surface respect to the BZ1. (g) Plot of the Maximally Localized Wannier Functions for the Ta $t_{2g}$ manifold, as observed from [110] direction.

In order to understand the effects of confinement along the [110] direction on the KTO bulk electronic structure we have used *BinPo*, an open-source code to compute the electronic properties of 2DEGs with high accuracy at low computational cost.[24] The free parameter in the calculation is the surface potential ($V_0$), a parameter related to the carrier density and therefore to the Fermi surface size. We have set $V_0$ so that the calculated band structure matches the experimentally determined Fermi wavevectors. The computed band structure along the high symmetry directions X-Γ-X and Y-Γ-Y is shown in Figures 3(a) and (b), respectively, with the calculated Fermi surface in Figure 3(c). Our calculations indicate that the maximum carrier density is reached at a depth of approximately 15 Å below the surface. The electronic structure calculation, projected onto the first 6 layers, reveals that this depth correspond mainly to the two bands with larger bandwidth, which is consistent with the bands detected in our experimental results.[42] The Fermi surface displays an evident anisotropy, with states having $d_{xy}$, $d_{yz}$ and $d_{zx}$ orbital character shown respectively in red, green and blue; the states appearing in cyan have a mixed $d_{yz}/d_{zx}$ character as indicated in the inset of Figure 3(c). The calculated orbital character accurately represents the experimental results, as LV (LH) photons excite $d_{yz}/d_{zx}$ ($d_{xy}$) electronic states describing the ellipse with a major axis along Γ-X (Γ-Y) (*cf.* Figures 2(c) and 2(d)). Taking into account the largest Fermi surface sheets that correspond to the two bands with largest band width in Figure 3(a), we estimate a carrier density $n_{2D} = 7 \times 10^{13}$ cm$^{-2}$, in agreement with transport experiments for superconducting samples.[6,9] The dispersions along X-Γ-X in Figures 2(a) and 2(b) also show strong polarization dependent modulation of intensity, providing insight into the energy dependence of the orbital character of the bands. By comparing the experimental dispersions to the computed band structure shown in Figure 3(a) we can validate the orbital character found in our calculation. The two broadest bands that appear with $k_F \sim 0.17$ and $0.15$ Å$^{-1}$, which have mixed $d_{yz}/d_{zx}$ orbital character according to our calculation, are detected with LV photons as shown in Figure 2(b). Conversely the region close to Γ point ($k_x = 0$), appears more intense when measured with LH photons as shown in Figure 2(a) due to the $d_{xy}$ preferential orbital character. The combination of the experiments and calculation allows us to clarify the effect of confinement along the [110] direction. The Γ-point degeneracy of the J = 3/2 bands observed in bulk KTO is broken by confinement along the

[110] direction and the lowest lying bands have mixed $d_{yz}$ and $d_{zx}$ orbital character due to their equivalence when viewed along the [-110] direction. The dispersion shown in Figure 3(b) along the Y-Γ-Y path is similar to the confined system along [001] with a ladder of intertwined heavy and light bands, however in the present case the bands have preferential $d_{xy}$ and $d_{yz}/d_{zx}$ orbital character, respectively.[30,31] There is a noticeably different situation along the X-Γ-X path, where an exact mix of $d_{zx}$ and $d_{yz}$ orbital character arises for the lowest lying subband, and a larger Rashba splitting is present.

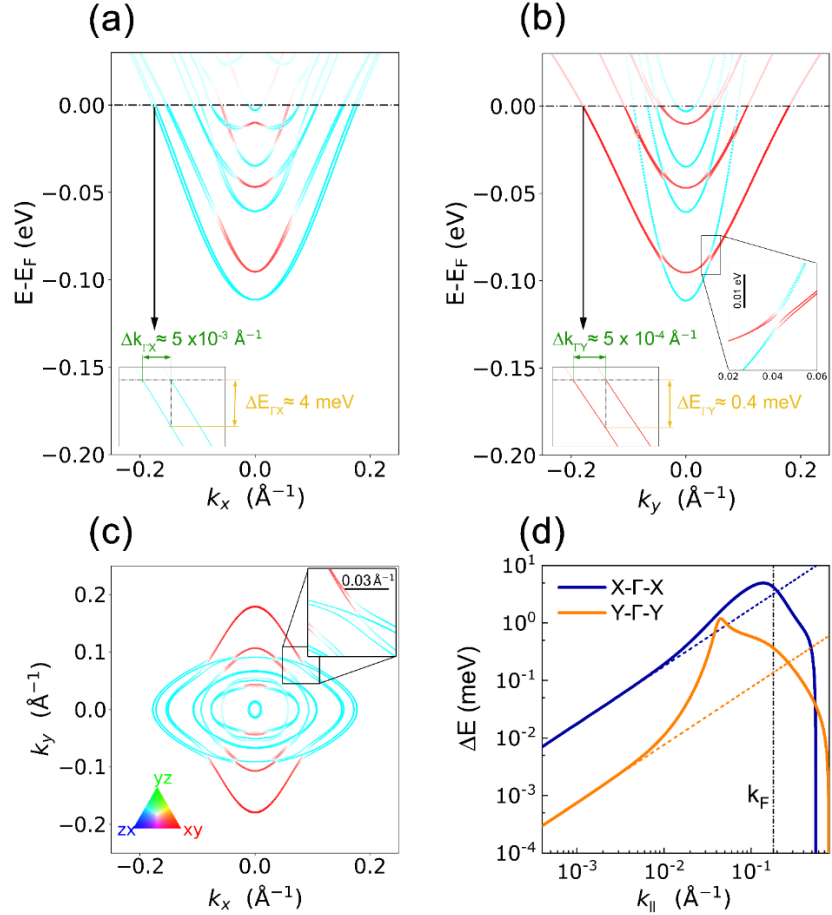

**Figure 3.** (a) Energy-momentum dispersion along the X-Γ-X path. Insert: Rashba energy and momentum spin-splitting at the Fermi level. (c) Energy-momentum dispersion along the Y-Γ-Y path. Inserts: Rashba energy and momentum spin-splitting at the Fermi level and at the avoided crossing point. (b) Orbitally projected Fermi surface. (d) Energy difference between spin-up and spin-down states (ΔE), along the X-Γ-X (blue) and Y-Γ-Y (orange) high-symmetry paths.

We now turn our attention to the strong Rashba splitting in the system expected from the combined breaking of inversion symmetry at the Al/KTO interface and the large atomic SO coupling of Ta. In the inset of Figure 3(a) we show that along the Γ-X path the momentum (energy) spin-splitting of the bands at $E_F$ reaches $\Delta k_{\Gamma X} \approx 5\cdot10^{-3}$ Å$^{-1}$ ($\Delta E_{\Gamma X} = 4$ meV) in the first subband, at least an order of magnitude larger than in other KTO-based 2DEGs [38]. On the other hand, the momentum (energy) spin-splitting along the Γ-Y path at $E_F$ is $\Delta k_{\Gamma Y} \approx 5\cdot10^{-4}$ Å$^{-1}$ ($\Delta E_{\Gamma Y} = 0.4$ meV) as observed in the inset of Figure 3(b), an order of magnitude lower than the former high symmetry direction and consistent with the anisotropy of the electronic structure. In conventional systems, the Rashba coefficient ($\alpha_R$) is defined as $\Delta E = 2\cdot\alpha_R\cdot k_{//}$, where ΔE is the energy difference between spin-up and spin-down states. In Figure 3(d) we show the dependence of ΔE with momentum along Γ-Y (orange) and Γ-X (blue), together with a linear fit for low $k_{//}$ values.[16] From these fits, we obtain Rashba constants of $α_R = 8·10^{-4}$ and $2·10^{-2}$ eV Å for the Γ-Y and Γ-X directions, respectively. However, the linear relation between energy splitting and momentum does not hold in KTO-based 2DEGs, as evident from Figure 3(d). Along the Γ-Y direction the splitting reaches a maximum at the crossing of the light (cyan) and heavy (red) bands shown in Figure 3(b) and inset. A similar behavior was previously observed in STO-based 2DEGs. [21] Remarkably, along the Γ-X direction we find an order of magnitude larger spin-splitting, which is particularly striking in the absence of band crossings. In order to explain this difference, arguments beyond the large atomic SOC of Ta and the electric field near the interface are necessary; the required contributions should affect differently the electronic structure along Γ-X and the Γ-Y direction. We found that the large anisotropic splitting can be rationalized based on the orbital angular momentum formulation for the Rashba effect.[49–51] This approach takes into account the orbital angular momentum (**L**) and the spin angular momentum (**S**), and it states that the magnitude and direction of both quantities are relevant, and an enhancement of **L**·**S** will lead to a large Rashba splitting. Along the Γ-X (Γ-Y) direction, **L** and **S** are parallel or antiparallel (perpendicular), thus maximizing (minimizing) the Rashba spin splitting. [42]

**Conclusion**

In summary, we have studied the electronic properties of the KTO(110)-2DEG formed after deposition of a thin Al layer on the clean substrate surface. Our ARPES measurements reveal electron-like bands with ~120 meV bandwidth and a strong dependence on the probing light polarization. The ARPES data allows us to determine that the Fermi surface is formed by at least two elliptical sheets, with their major axes aligned with the [001] and [-110] crystallographic directions and preferential $d_{xy}$ and $d_{zx}/d_{yz}$ orbital character, respectively. This anisotropic orbital character results from the symmetry of the 5$d$ orbitals with respect to the confinement direction. We used tight-binding slab calculations as implemented in BinPo to determine the band structure of the system. The resulting electronic structure is in fair agreement with the experiment. A closer inspection of the calculation reveals the presence of an unconventional and anisotropic Rashba effect in the system. The Rashba splitting at the Fermi level for the first subband is $ΔE_{ΓX} ≈ 4$ meV along the Γ-X direction, the largest among STO- and KTO-2DEGs in different orientations and with a corresponding momentum splitting of $Δk_{ΓX} ≈ 5·10^{-3}$ Å$^{-1}$ that could, in principle, be resolved in an ARPES experiment. In our experiment, we observed a linewidth of at least an order of magnitude larger than $Δk_{ΓX}$, suggesting that we are limited by sample quality. Remarkably, the large splitting is found in a band that presents a $d_{zx}/d_{yz}$ orbital character mix, in the absence of so-called avoided crossings. However, an enhanced interplay between orbital and spin angular momentum is observed where large spin-splitting is found, this highlights the importance of taking into account this interaction in the study of the Rashba effect in oxide-based 2DEGs. The electronic structure presented in this manuscript is a useful platform for the interpretation of future experiments in spin-orbitronics and superconducting effects on KTO-2DEGs.

**Acknowledgments**

We thank Siobhan Mckeown Walker for valuable discussions. We acknowledge Jordi Prat for technical support during ARPES experiments. This work has been supported by Comunidad de Madrid (Atracción de Talento grant No. 2018-T1/IND-10521 and 2022-5A/IND-24230) and by MICINN PID2019-105238GA-I00. We acknowledge allocation of measurement time at ALBA under proposals 2021024953 and 2022025693. LOREA was co-funded by the European Regional Development Fund (ERDF) within the Framework of the Smart Growth Operative Programme 2014-2020.

**Conflict of Interest**

The authors declare no conflict of interest.

**Data Availability Statement**

The data that support the findings of this study are available from the corresponding author upon reasonable request.

**Experimental**

**ARPES**

ARPES experiments were performed at the LOREA beamline of the ALBA synchrotron (Spain) using an MBS A1 hemispherical analyzer with horizontal slit. The photon energy used was in the 50-110 eV range and sample temperature was T = 15 K.

**Calculations**

The calculations were performed using *BinPo* code, a tool for computing the electronic properties of 2DEGs. [24] A slab tight binding Hamiltonian is constructed from relativistic density functional theory calculations represented in the basis of maximally localized Wannier functions. [48] Initially, a linear potential energy is added to the slab Hamiltonian accounting for the band-bending potential. By solving the Schrödinger-Poisson scheme along the slab, the self-consistent potential is then found. [52] Subsequently, the orbitally projected band structure and Fermi surface were computed. We have taken 40 Ta planes stacked along the [110] direction of cubic $KTaO_3$. We have tuned the value of $V_0$ to approximately match our Fermi wavevectors and the estimated bandwidths. Importantly, when solving the Poisson equation, we have considered an electric field-dependent relative permittivity model for $KTaO_3$. Detailed information about the relative permittivity model as well as all the parameters of the calculations can be found in the Supporting Information.[42]